\documentclass[prl, showkeys, amsmath,amssymb,aps,twocolumn, superscriptaddress, nobibnotes, longbibliography]{revtex4-2}
\usepackage{amsfonts}
\usepackage{graphicx} 
\usepackage{physics}
\usepackage{xcolor}

\usepackage{xcolor}
\newif\ifshowchanges
 \showchangesfalse  

\DeclareRobustCommand{\rev}[1]{%
    {\ifshowchanges\color{blue}\fi#1}%
}

\newenvironment{revision}
    {\begingroup\ifshowchanges\color{blue}\fi}
    {\endgroup}

\begin{document}
\title{Double quantum spin Hall phase in bilayer ZrTe$_5$}

\author{Chao Chen Ye}
\thanks{c.chen.ye@rug.nl}
\affiliation{Zernike Institute for Advanced Materials, University of Groningen, Nijenborgh 3, 9747 AG Groningen, The Netherlands}
\author{G{\'a}bor Kalla}
\affiliation{Zernike Institute for Advanced Materials, University of Groningen, Nijenborgh 3, 9747 AG Groningen, The Netherlands}
\author{Marcio Costa}
\email{mjtcosta@id.uff.br}
\affiliation{Instituto de Física, Universidade Federal Fluminense, Niterói, Rio de Janeiro, Brazil}
\author{Jianting Ye}
\affiliation{Zernike Institute for Advanced Materials, University of Groningen, Nijenborgh 3, 9747 AG Groningen, The Netherlands}
\author{Jagoda S{\l}awi{\'n}ska}
\thanks{jagoda.slawinska@rug.nl}
\affiliation{Zernike Institute for Advanced Materials, University of Groningen, Nijenborgh 3, 9747 AG Groningen, The Netherlands}


\begin{abstract}
\textbf{\hspace{-2.5ex}Abstract\linebreak}
Quantum spin Hall (QSH) insulators host helical edge states that enable dissipationless transport, but realizing multiple conducting channels in experimentally accessible materials with large band gaps remains challenging. Here, we show that AB-stacked untwisted bilayer ZrTe$_5$ naturally realizes a double QSH phase in its energetically most stable structure, revealing an unexpected multichannel transport regime. The phase hosts two pairs of helical edge states and a doubled quantized spin Hall conductivity, with spin Chern number $C_s=2$ despite the conventional $\mathbb{Z}_2=0$ classification. Lattice deformation tunes the system through a metallic state to a conventional QSH phase with $\mathbb{Z}_2=1$, enabling control of the number of helical channels. Remarkably, the quantized spin Hall response persists over an energy window substantially larger than the bulk band gap, exceeding 100~meV. Our results establish bilayer ZrTe$_5$ as an experimentally accessible multichannel QSH system in which interlayer coupling and lattice deformation control topological edge transport.
\end{abstract}
\maketitle

Quantum spin Hall (QSH) insulators are two-dimensional (2D) topological phases that host conducting edge states protected from backscattering by time-reversal symmetry~\cite{qi_2011,kane_mele_2005,bernevig_2006,konig_2007, qian_2014}. In conventional QSH insulators characterized by the topological invariant $\mathbb{Z}_2 = 1$, an odd number of pairs of helical edge states gives rise to quantized edge transport. Systems with an even number of pairs of helical edge states, referred to as double QSH insulators, have $\mathbb{Z}_2 = 0$ and can be characterized by an even spin Chern number ($C_s$)~\cite{sheng_2006,prodan_2009,ezawa_2013,bai_2022,liu_2024}.
Such phases have been theoretically predicted to exhibit increased edge conductance and robustness against magnetic perturbations~\cite{sheng_2005,prodan_2009,yang_2011,ezawa_2013}. 
However, realizing these properties experimentally remains challenging: double QSH phases have only recently been reported in twisted van der Waals (vdW) materials~\cite{Kang2024,kang_2024}, while experimentally accessible realizations with sizeable band gaps suitable for high-temperature operation remain extremely rare~\cite{roadmap_2024}.

Zirconium pentatelluride (ZrTe$_5$) is a layered vdW material that has been extensively studied for its tunable topological properties~\cite{weng_2014,manzoni_2016,chen_2017,tang_2019,galeski_2021,tang_2021,wang_2021,chenye_2025}. The three-dimensional (3D) crystal lies in close proximity to a topological phase transition and can be tuned via external strain~\cite{fan_2017,mutch_2019,xu_2018,zhang_2021,tajkov_2022,chenye_2025}, but its sensitivity to lattice deformations along three crystallographic directions, together with the presence of a large number of conducting channels, limits its suitability for controlled edge transport. Reducing the thickness to a few layers could provide greater control over electronic and topological behaviour, but only a few experiments have achieved this regime~\cite{qiu_2016, tang_2021}. In the 2D limit, monolayer ZrTe$_5$ has long been predicted to realize a QSH phase with a large band gap~\cite{weng_2014}, yet its experimental realization remains challenging despite recent efforts~\cite{xu_2024}. Bilayer ZrTe$_5$, which is more accessible via exfoliation, provides a natural intermediate system that remains comparatively unexplored, combining experimental feasibility with tunable topological behaviour~\cite{tajkov_2023}.

Here, based on first-principles calculations, we show that bilayer ZrTe$_5$ hosts a tunable topological phase diagram under lattice deformations, comprising gapped phases with $\mathbb{Z}_2 = 0$ and $\mathbb{Z}_2 = 1$ separated by a metallic state. The $\mathbb{Z}_2 = 0$ phase, corresponding to the most stable structure, realizes a double QSH state characterized by even spin Chern number ($C_s=2$), while the $\mathbb{Z}_2 = 1$ or $C_s=1$ phase corresponds to a conventional QSH insulator, with a transition between them driven by uniaxial strain. The double QSH phase exhibits two pairs of helical edge states and a corresponding double quantized spin Hall conductivity (SHC). Both conventional and double QSH phases exhibit wide quantized SHC plateaus, indicating a large effective transport gap. These results demonstrate that multi-channel QSH phases under spin Chern number classification can arise in realistic vdW materials without twisting and establish bilayer ZrTe$_5$ as a realistic platform for strain-tunable transitions between conventional and double QSH insulators and robust high-temperature topological transport~\cite{prodan_2009,yang_2011,ezawa_2013,Kang2024}.

\begin{figure*} 
	\centering
	\includegraphics[width=\textwidth]{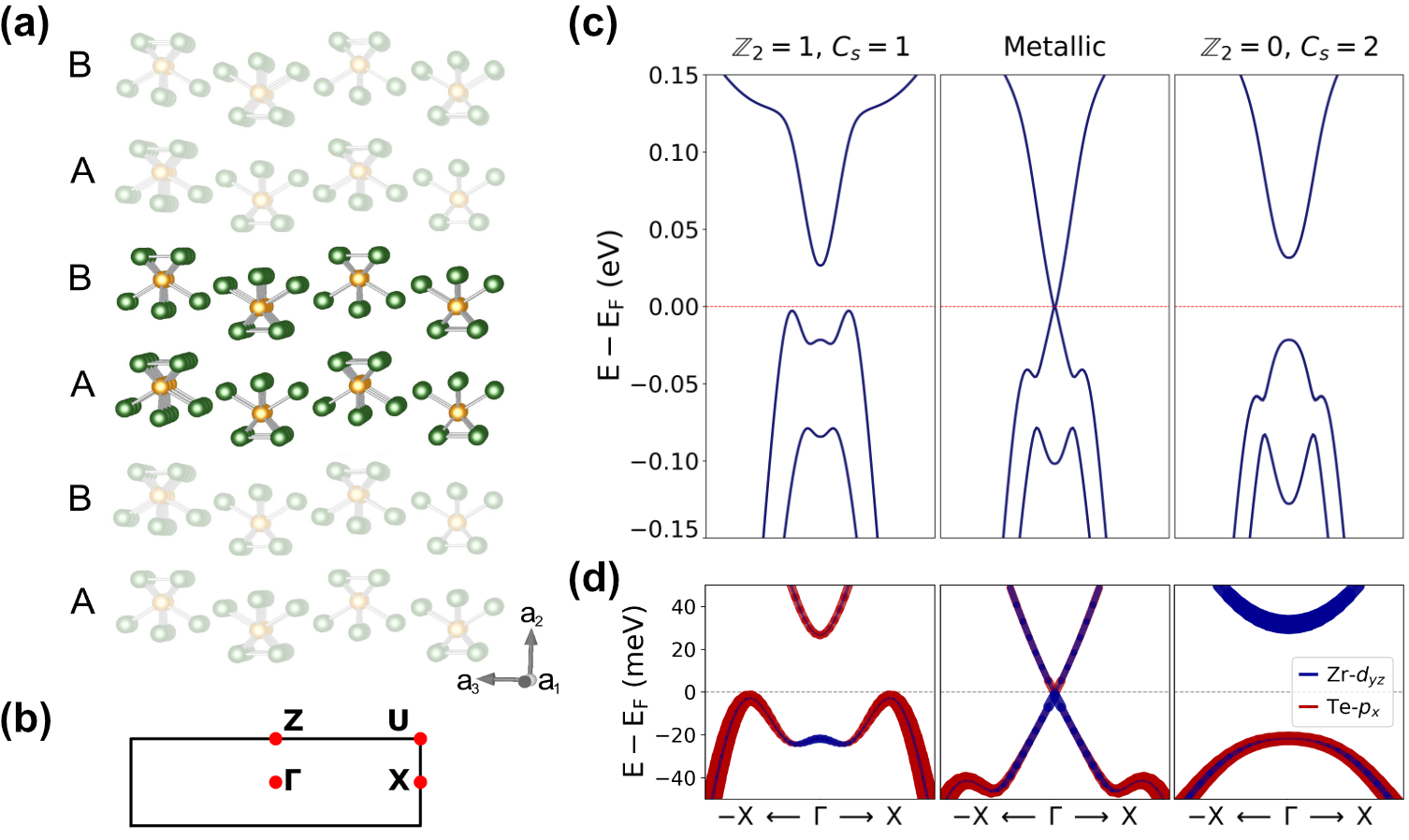}
	\caption{
		(\textbf{a}) Crystal structure of 3D and bilayer ZrTe$_5$; the bilayer that can be exfoliated from the 3D crystal is highlighted. (\textbf{b}) The 2D Brillouin zone. (\textbf{c}) Bandgap evolution along the $-X - \Gamma - X$ $k$-path illustrating the topological phase transition. (\textbf{d}) Orbital-projected band inversion during the topological phase transition.}
	\label{fig:crystal_bandgap_evolution}
\end{figure*}

The crystal structure of 3D ZrTe$_5$ contains two vdW layers in AB stacking within an orthorhombic Bravais lattice~\cite{weng_2014,chenye_2025}, as depicted in Fig.~\ref{fig:crystal_bandgap_evolution}(\textbf{a}). Exfoliating along the stacking direction, $y \parallel \mathbf{a}_2$, yields a single 2D AB-stacked bilayer arrangement, while further exfoliation leads to the monolayer structure. However, monolayer flakes are significantly less abundant than few-layer (including bilayer) flakes in standard exfoliation methods~\cite{hernandez_2008, coleman_2011, backes_2020, wang_exfoliation_2023}. Therefore, we focus on the bilayer system in the main text, while the monolayer calculations are reported in the Supporting Information (Section I).

The topological properties of ZrTe$_5$ are tunable by external strain, which can be simulated by adiabatic lattice deformation in DFT calculations~\cite{fan_2017,manzoni_2016,mutch_2019,xu_2018,zhang_2021,tajkov_2022,chenye_2025}. In the following, we consider biaxial lattice deformations and analyze the material for distinct values of the lattice parameters $\mathbf{a}_1$ and $\mathbf{a}_3$. The first topological feature we find is the bandgap closing and reopening shown in Fig.~\ref{fig:crystal_bandgap_evolution}(\textbf{c}), which is considered a signature of a topological phase transition~\cite{bernevig_2006,mutch_2019,tajkov_2023}. It is accompanied by the characteristic band inversion between the valence and conduction bands near the $\Gamma$ point, where the topological phase transition occurs~\cite{weng_2014,chen_2017,wang_2021}. A representative evolution of the orbital character is shown in Fig.~\ref{fig:crystal_bandgap_evolution}(\textbf{d}), where the band inversion can be clearly seen between the Te-$p_x$ and Zr-$d_{yz}$ orbitals near $\Gamma$. In the $\mathbb{Z}_2=0$ phase, the valence band is dominated by the Te-$p_x$ orbital, while the conduction band is mainly of Zr-$d_{yz}$ character; after the topological phase transition to the regime with $\mathbb{Z}_2=1$, i.e., after crossing the metallic phase, the orbital characters of the valence and conduction bands are inverted.

\begin{figure*} 
	\centering
	\includegraphics[width=\textwidth]{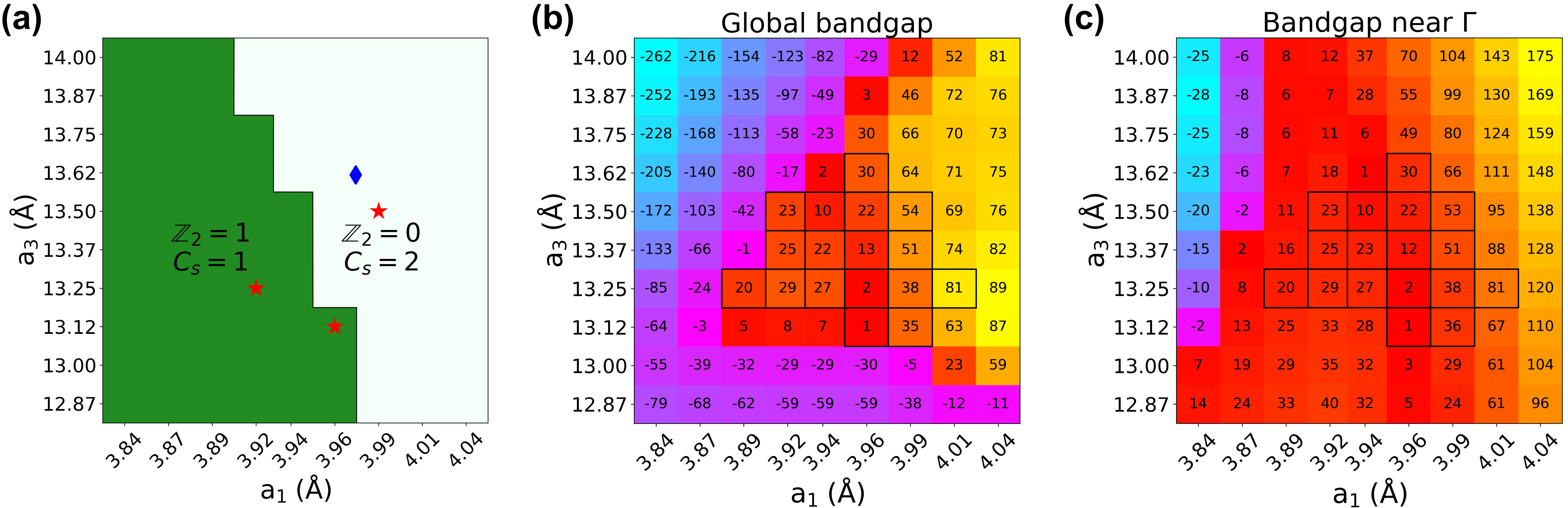}
	\caption{
		(\textbf{a}) Topological phase diagram of bilayer ZrTe$_5$ under lattice deformation, where $\mathbb{Z}_2 = 1$ or $C_s=1$ indicates the conventional QSH phase and $\mathbb{Z}_2 = 0$ denotes a distinct topological regime corresponding to the double QSH phase characterized by $C_s=2$. The three red stars mark the deformations of the three phases shown in Figs.~\ref{fig:crystal_bandgap_evolution}(\textbf{c}) and~\ref{fig:crystal_bandgap_evolution}(\textbf{d}). The dark diamond denotes the fully relaxed DFT structure. (\textbf{b}) The global bandgap and (\textbf{c}) the local bandgap near $\Gamma$ for each set of lattice parameters, in units of meV. The black frames highlight the deformations for which the global bandgap is located near $\Gamma$, where the bandgap closing and reopening occur.}
	\label{fig:z2}
\end{figure*}

The topological invariant is considered the key quantity for determining the topological properties of a material and the most rigorous way to confirm topological phases. Figure~\ref{fig:z2}(\textbf{a}) shows the resulting topological phase diagram, based on both the $\mathbb{Z}_2$ and $C_s$ invariants, under biaxial lattice deformation as a function of ${a}_1$ and ${a}_3$. Here, $\mathbb{Z}_2 = 1$ or $C_s = 1$ indicates the QSH phase, while $\mathbb{Z}_2 = 0$ corresponds to a double QSH phase that is not captured by the $\mathbb{Z}_2$ classification but instead by the even spin Chern number  $C_s=2$. The topological phase transition occurs at the boundary between these two regimes, mainly distinguished by the value of ${a}_1$, which is consistent with the bandgap closing and reopening in Fig.~\ref{fig:crystal_bandgap_evolution}(\textbf{c}) and in the Supporting Information (Section II). \rev{A strain-induced transition between phases with $\mathbb{Z}_2=1$ and $\mathbb{Z}_2=0$ was also reported in Ref.~\cite{tajkov_2023},  with qualitatively consistent results regarding the evolution of the $\mathbb{Z}_2$ invariant. However, the $\mathbb{Z}_2=0$ phase was previously interpreted as topologically trivial. By calculating the spin Chern number, we establish that this phase has $C_s=2$ and therefore supports a double QSH state.} In the diagram, the red stars mark the three deformations at which the band evolution is presented in Fig.~\ref{fig:crystal_bandgap_evolution}(\textbf{c}, \textbf{d}). Remarkably, the most stable structure of bilayer ZrTe$_5$, marked by the dark diamond, is located close to the topological phase transition region. This proximity implies easy tunability of the topological properties by external strain and may also explain possible experimental discrepancies arising from sample quality~\cite{mutch_2019,zhang_2021,tajkov_2022,chenye_2025}.

\begin{revision}
Surprisingly, the topological invariants remain relatively insensitive to the strength of interlayer interaction, and external perturbations, such as an applied electric field. Our additional calculations performed for a representative configuration with $C_s=2$ demonstrate that large electric fields are needed to drive a transition to conventional QSH, while artificial tuning of the interlayer interaction further reveals the robustness of the double QSH phase (see Supporting Information, Section III). 
\end{revision}

\begin{figure*} 
	\centering
	\includegraphics[width=\textwidth]{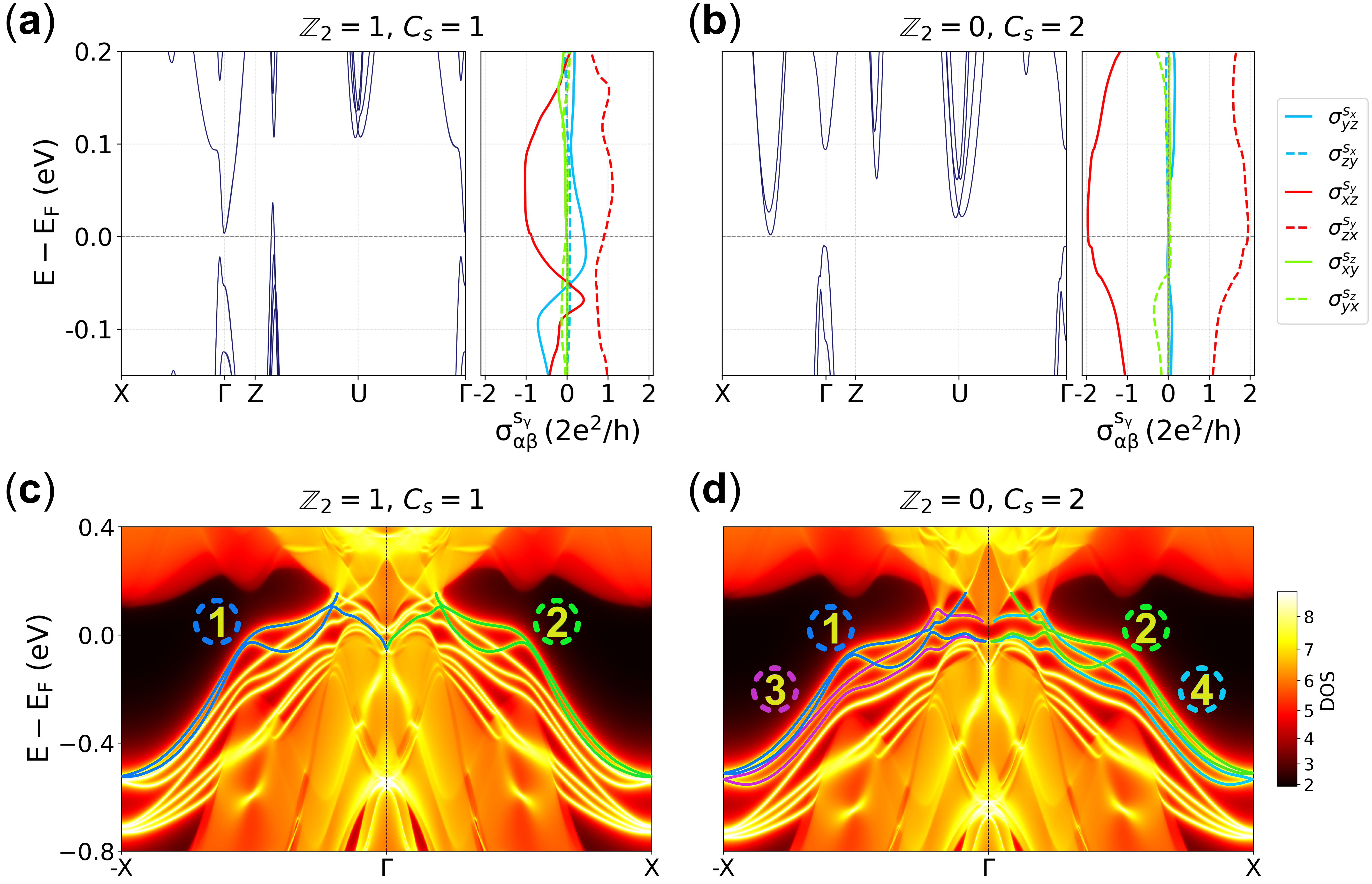}
	\caption{
		Bulk band structures and the corresponding SHC for the representative lattice deformations characterized by (\textbf{a}) $\mathbb{Z}_2=1$, (\textbf{b}) $\mathbb{Z}_2=0$. (\textbf{c}) $\mathbb{Z}_2=1$ regime has two robust edge states connecting valence and conduction bands, as numbered and colored. (\textbf{d}) $\mathbb{Z}_2=0$ phase has four robust edge states connecting valence and conduction bands, as numbered and colored.}
	\label{fig:shc_bands_bulk_edge_l}
\end{figure*}

The value of the bandgap is essential for applications of topological materials and is a key factor that constrains the operating temperature of the QSH effect~\cite{qian_2014,xu_2024,roadmap_2024}. The global bandgap and the local one near $\Gamma$, calculated within the Perdew-Burke-Ernzerhof exchange-correlation functional~\cite{perdew_1996}, are shown in Figs.~\ref{fig:z2}(\textbf{b, c}), respectively. Importantly, bandgap closing can occur only at the $\Gamma$ point, while the valence band maximum and conduction band minimum may lie away from this point. The relevant deformations are those for which the global bandgap is located near $\Gamma$, highlighted by the black frames in Figs.~\ref{fig:z2}(\textbf{b, c}). Here, $29$~meV and $81$~meV are the largest bandgaps found for the $\mathbb{Z}_2=1$ and $\mathbb{Z}_2=0$ regimes, respectively. An additional calculation using the Heyd-Scuseria-Ernzerhof (HSE) functional~\cite{hse_2009} increases the value of the former to $\sim 50$~meV, suggesting that the bandgaps in Figs.~\ref{fig:z2}(\textbf{b, c}) can be nearly twice as large. Bands located at $k$ points other than $\Gamma$ can lie closer to the Fermi level. The deformations for which the conduction band edge lies below the valence band edge are indicated by negative bandgap values in Figs.~\ref{fig:z2}(\textbf{b, c}).

We next reveal how the different topological phases translate into quantum spin Hall response by calculating the spin Hall conductivity (SHC) for all considered deformations. Figure~\ref{fig:shc_bands_bulk_edge_l}(\textbf{a}, \textbf{b}) shows representative results together with the corresponding full $k$-path band structures, where Fig.~\ref{fig:shc_bands_bulk_edge_l}(\textbf{a}) corresponds to the $\mathbb{Z}_2 = 1$ regime and Fig.~\ref{fig:shc_bands_bulk_edge_l}(\textbf{b}) to the $\mathbb{Z}_2 = 0$ phase. Among the six symmetry-allowed SHC tensor components~\cite{roy_2022}, $\sigma^{s_y}_{ x z}$ and $\sigma^{s_y}_{z x}$ dominate, giving rise to spin currents with the out-of-plane spin polarization. The components $\sigma^{s_x}_{  y z}$ and $\sigma^{s_z}_{  y x}$ are small for most deformations but can reach up to half of the quantized value in a few cases, while $\sigma^{s_x}_{ z y}$ and $\sigma^{s_z}_{ x y}$ remain negligible for all deformations.

Quantized SHC is observed only for the dominant components $\sigma^{s_y}_{xz}$ and $\sigma^{s_y}_{zx}$. In the $\mathbb{Z}_2 = 1$ regime, we obtain $\sigma^{s_y}_{xz} = 2 \, e^2/h$ [Fig.~\ref{fig:shc_bands_bulk_edge_l}(\textbf{a})], consistent with a conventional QSH phase. In contrast, in the $\mathbb{Z}_2 = 0$ regime, the SHC reaches $\sigma^{s_y}_{xz} = \sigma^{s_y}_{zx} = 2 \, (2 \, e^2/h)$ [Fig.~\ref{fig:shc_bands_bulk_edge_l}(\textbf{b})], indicating a doubled quantized response. While the quantization in the $\mathbb{Z}_2 = 0$ regime appears inconsistent with the conventional classification, the $\mathbb{Z}_2$ invariant defines the modulo of the number of pairs of helical edge states, and systems with an even number of pairs can still host robust edge states characterized by $C_s$ invariant~\cite{sheng_2006,bai_2022,liu_2024,tan_2025}. 
For comparison, we show in the Supporting Information (Section I) that monolayer ZrTe$_5$ remains a conventional QSH insulator with quantized SHC of $2 \, e^2/h$ across all deformations. Since the bilayer consists of mirror plane AB stacking of two monolayer QSH insulators, the interaction between opposite conducting helical edge channels with the same spin orientation can induce hybridization~\cite{stuhler_2022}. Therefore, a pair of conduction channels in the $\mathbb{Z}_2 = 1$ or $C_s=1$ regime is (partially) lifted without breaking time-reversal symmetry. In contrast, in the $C_s=2$ phase, the two pairs of helical edge states coexist without hybridization, yielding a double quantized SHC. 

We further analyze the relation between the SHC and the bulk band structure. Figure~\ref{fig:shc_bands_bulk_edge_l}(\textbf{a}, \textbf{b}) shows that many bands do not contribute significantly to the quantized SHC. Notably, the SHC plateaus can extend well beyond the energy window defined by the bandgap near $\Gamma$. For instance, in Fig.~\ref{fig:shc_bands_bulk_edge_l}(\textbf{a}), additional states present along the $Z-U$ direction contribute to subleading components such as $\sigma^{s_x}_{yz}$ but do not affect the quantized values of $\sigma^{s_y}_{xz}$ and $\sigma^{s_y}_{zx}$, 
Similarly, in Fig.~\ref{fig:shc_bands_bulk_edge_l}(\textbf{b}), the extended SHC plateau remains robust despite the presence of additional conduction bands away from the $\Gamma$ point. These observations indicate that the effective transport gap relevant for the quantized response can be substantially larger than the global bandgap or the local gap at $\Gamma$.

The origin of the quantized SHC plateaus and their relation to the band structure can be understood from the edge state calculations shown in Fig.~\ref{fig:shc_bands_bulk_edge_l}(\textbf{c}, \textbf{d}). The spectra reveal helical edge states connecting the valence and conduction bands, with one pair in the $\mathbb{Z}_2 = 1$ or $C_s=1$ regime and two pairs in the $\mathbb{Z}_2 = 0$ or $C_s=2$ phase, resulting either in one or two pairs of conductance channels, respectively. Other edge states are also present but are not topologically protected. Notably, the helical edge states extend deep into both the valence and conduction parts of the spectrum, which explains the extended SHC plateaus. As a result, the quantized response remains robust even in the presence of additional bulk bands.
Consequently, while the bulk bandgaps are on the order of a few tens of meV, the effective transport gap associated with the quantized SHC reaches values approaching 100~meV, and in some cases even 200~meV (see Supporting Information), and remains largely unaffected by additional bulk states or nonhelical edge states. \rev{The explicit calculations for a nanoribbon distinguishing edge and bulk contributions to transport further support these claims (see Supplementary Information, Section IV).}

In summary, we show that AB-stacked bilayer ZrTe$_5$ realizes a tunable topological phase diagram with both conventional and double QSH phases, originating from the interlayer interaction between two monolayer QSH insulators. The fully relaxed structure lies close to the phase boundary, enabling efficient control via moderate strain. In addition to the $\mathbb{Z}_2 = 1$ or $C_s=1$ phase with a quantized spin Hall conductivity of $2 \, e^2/h$, the system exhibits a $\mathbb{Z}_2 = 0$ or $C_s=2$ regime characterized by a doubled quantized response, $2 \, (2 \, e^2/h)$, arising from two pairs of helical edge states. 
More broadly, our results suggest that double quantum spin Hall phases may be more widespread than previously recognized and could have been overlooked in systems where topology is assessed solely through the $\mathbb{Z}_2$ classification. We demonstrate that multi-channel QSH phases can emerge as ground states in realistic vdW bilayers without twisting and can be tuned within a single system that highly increase the signal-to-noise ratio, providing a practical route for topological materials discovery suitable for 2D QSH devices. 

\textbf{Methods} \par
The first-principles calculations based on DFT are performed using the \texttt{Vienna Ab initio Simulation Package} (VASP)~\cite{vasp_1993, vasp_1996, kresse_1996_2}. The same input parameters are used for each lattice deformation to ensure consistent and comparable results. The projector-augmented-wave method is used to describe the ion-electron interaction~\cite{blochl_1994,kresse_joubert_1999}. The kinetic-energy cutoff for the plane-wave basis is set to 350~eV, and energy convergence threshold in self-consistent calculations to $10^{-6}$~eV. The generalized gradient approximation of Perdew, Burke, and Ernzerhof is used for the exchange-correlation functional~\cite{perdew_1996}. \rev{The structures were relaxed for each fixed lattice parameters, leaving the atomic positions as the only degree of freedom. Van der Waals interactions were taken into account using DFT-D3 method~\cite{dft-d3}.} Spin-orbit coupling is taken into account in all calculations except for the relaxations, and the symmetries are kept switched on to preserve the topological properties. The Brillouin zone is sampled using a $20 \times 1 \times 6$ $k$-point mesh centered at $\Gamma$ for atomic relaxation and self-consistent calculations, and increased to a $26 \times 1 \times 8$ mesh for non-self-consistent calculations. 

In the post-processing step, we used \texttt{Wannier90}~\cite{mostofi_2008,pizzi2020} to build the Wannier tight-binding Hamiltonian by projecting all Te[s, p] and Zr[s, p, d] atomic orbitals. The final Hamiltonian is used to calculate the 2D topological $\mathbb{Z}_2$ index and the WCC evolution diagrams in both \texttt{Z2Pack}~\cite{z2pack_2011,z2pack_2017} and \texttt{WannierTools}~\cite{wanniertools_2018}. The latter program was also used to compute edge states using the semi-infinite Green's function approach~\cite{lopez_sancho_1985}. The global and local bandgaps are calculated from both the VASP and Wannier90 smooth band structures, where the value is defined as the energy difference between the conduction- and valence band edges. The SHC and $C_s$ is calculated based on pseudo-atomic orbitals (PAO) tight-binding Hamiltonians using \texttt{PAOFLOW}~\cite{paoflow_2018,paoflow_2021, paoflow_2026} with an ultra-dense k-mesh of $100 \times 1 \times 30$. The space group is determined by \texttt{Spglib}~\cite{Togo31122024} and \texttt{IrRep}~\cite{irreps}. To check for possible band inversion after bandgap reopening, we calculated irreducible representations at $\Gamma$ for both the valence and conduction bands using \texttt{IrRep}, as well as orbital-projected band structures from VASP and spin-projected band structures from \texttt{WannierTools}. 

\medskip
\vskip 1.0 em
\textbf{Supporting Information} \par
\rev{Calculations of $\mathbb{Z}_2$ invariants, band structures and spin Hall conductivity for representative configurations of monolayer and bilayer ZrTe$_5$, projected spin spectrum, additional calculations demonstrating the robustness of double quantum spin Hall phase, transport calculations for a nanoribbon, and symmetry analysis.}

\medskip
\textbf{Data availability} \par
The data that support the findings of this study are openly available at https://doi.org/10.34894/BAKZFU. 

\medskip
\textbf{Acknowledgements} \par
We thank M. Mostovoy, Y. Kreminska and L. Eek for helpful discussions. We acknowledge the research program “Materials for the Quantum Age” (QuMat) for financial support. This program (registration number 024.005.006) is part of the Gravitation program financed by the Dutch Ministry of Education, Culture and Science (OCW). M.C. acknowledges the financial support of CNPq Grant 317320/2021-1, FAPERJ Grant E26/200.240/2023, INCT Materials Informatics and the computational resources of the Centro Multiusuário de Computação de Alto Desempenho (CEMUCAD-RJ). The calculations were carried out on the Dutch national e-infrastructure with the support of SURF Cooperative (EINF-10786) and on the Hábrók high-performance computing cluster of the University of Groningen.


\medskip

\bibliography{bib}

@article{vasp_1993,
	title = {Ab initio molecular dynamics for liquid metals},
	author = {Kresse, G. and Hafner, J.},
	journal = {Phys. Rev. B},
	volume = {47},
	issue = {1},
	pages = {558--561},
	numpages = {0},
	year = {1993},
	month = {Jan},
	publisher = {American Physical Society},
}

@article{vasp_1996,
	title = {Efficient iterative schemes for ab initio total-energy calculations using a plane-wave basis set},
	author = {Kresse, G. and Furthm\"uller, J.},
	journal = {Phys. Rev. B},
	volume = {54},
	issue = {16},
	pages = {11169--11186},
	numpages = {0},
	year = {1996},
	month = {Oct},
	publisher = {American Physical Society},
}

@article{kresse_1996_2,
	title = {Efficiency of ab-initio total energy calculations for metals and semiconductors using a plane-wave basis set},
	journal = {Comp. Mat. Sci.},
	volume = {6},
	number = {1},
	pages = {15-50},
	year = {1996},
	author = {G. Kresse and J. Furthmüller},
}

@article{perdew_1996,
	title = "{Generalized gradient approximation made simple}",
	author = {Perdew, J. P. and Burke, K. and Ernzerhof, M.},
	journal = {Phys. Rev. Lett.},
	volume = {77},
	issue = {18},
	pages = {3865--3868},
	numpages = {0},
	year = {1996},
	month = {Oct},
	publisher = {American Physical Society},
}

@article{blochl_1994,
	title = {Projector augmented-wave method},
	author = {Bl{\"o}chl, Peter E.},
	journal = {Phys. Rev. B},
	volume = {50},
	issue = {24},
	pages = {17953--17979},
	year = {1994},
	month = {Dec},
	publisher = {American Physical Society},
	doi = {10.1103/PhysRevB.50.17953},
}

@article{kresse_joubert_1999,
	title = {From ultrasoft pseudopotentials to the projector augmented-wave method},
	author = {Kresse, Georg and Joubert, D.},
	journal = {Phys. Rev. B},
	volume = {59},
	issue = {3},
	pages = {1758--1775},
	year = {1999},
	month = {Jan},
	publisher = {American Physical Society},
	doi = {10.1103/PhysRevB.59.1758},
}

@article{mostofi_2008,
	title = {{wannier90: A tool for obtaining maximally-localised Wannier functions}},
	author = {Mostofi, Arash A. and Yates, Jonathan R. and Lee, Young-Su and Souza, Ivo and Vanderbilt, David and Marzari, Nicola},
	journal = {Comput. Phys. Commun.},
	volume = {178},
	number = {9},
	pages = {685--699},
	year = {2008},
	doi = {10.1016/j.cpc.2007.11.016},
}

@article{lopez_sancho_1985,
	title = {{Highly convergent schemes for the calculation of bulk and surface Green functions}},
	author = {{L{\'o}pez Sancho}, M. P. and {L{\'o}pez Sancho}, J. M. and {L{\'o}pez Sancho}, J. M. L. and Rubio, J.},
	journal = {J. Phys. F: Met. Phys.},
	volume = {15},
	number = {4},
	pages = {851--858},
	year = {1985},
	doi = {10.1088/0305-4608/15/4/009},
}

@article{pizzi2020,
	year = 2020,
	month = {jan},
	publisher = {{IOP} Publishing},
	volume = {32},
	number = {16},
	pages = {165902},
	author = {Giovanni Pizzi and Valerio Vitale and Ryotaro Arita and Stefan Bl{\"u}gel and Frank Freimuth and Guillaume G{\'e}ranton and Marco Gibertini and Dominik Gresch and Charles Johnson and Takashi Koretsune and Julen Iba{\~n}ez-Azpiroz and Hyungjun Lee and Jae-Mo Lihm and Daniel Marchand and Antimo Marrazzo and Yuriy Mokrousov and Jamal I. Mustafa and Yoshiro Nohara and Yusuke Nomura and Lorenzo Paulatto and Samuel Ponc{\'e} and Thomas Ponweiser and Junfeng Qiao and Florian Th{\"o}le and Stepan S. Tsirkin and Ma{\l}gorzata Wierzbowska and Nicola Marzari and David Vanderbilt and Ivo Souza and Arash A. Mostofi and Jonathan R. Yates},
	title = {Wannier90 as a community code: new features and applications},
	journal = {J. Phys. Condens. Matter},
	doi = {10.1088/1361-648X/ab51ff},
}

@article{Togo31122024,
	author = {Atsushi Togo and Kohei Shinohara and Isao Tanaka},
	title = {Spglib: a software library for crystal symmetry search},
	journal = {Sci. Technol. Adv. Mater: Meth.},
	volume = {4},
	number = {1},
	pages = {2384822},
	year = {2024},
	publisher = {Taylor \& Francis},
}

@article{irreps,
	title = {{IrRep: Symmetry eigenvalues and irreducible representations of ab initio band structures}},
	journal = {Comput. Phys. Commun.},
	volume = {272},
	pages = {108226},
	year = {2022},
	issn = {0010-4655},
	doi = {10.1016/j.cpc.2021.108226},
	author = {Mikel Iraola and Juan L. Mañes and Barry Bradlyn and Matthew K. Horton and Titus Neupert and Maia G. Vergniory and Stepan S. Tsirkin},
}

@article{paoflow_2018,
	title = "{PAOFLOW: A utility to construct and operate on ab initio Hamiltonians from the projections of electronic wavefunctions on atomic orbital bases, including characterization of topological materials}",
	journal = {Comp. Mat. Sci.},
	volume = {143},
	pages = {462-472},
	year = {2018},
	author = {Marco {Buongiorno Nardelli} and Frank T. Cerasoli and Marcio Costa and Stefano Curtarolo and Riccardo {De Gennaro} and Marco Fornari and Laalitha Liyanage and Andrew R. Supka and Haihang Wang}
}

@article{paoflow_2021,
	title = "{Advanced modeling of materials with PAOFLOW 2.0: new features and software design}",
	journal = {Comp. Mat. Sci.},
	volume = {200},
	pages = {110828},
	year = {2021},
	author = {Frank T. Cerasoli and Andrew R. Supka and Anooja Jayaraj and Marcio Costa and Ilaria Siloi and Jagoda S{\l}awi{\'n}ska and Stefano Curtarolo and Marco Fornari and Davide Ceresoli and Marco {Buongiorno Nardelli}}
}

@article{paoflow_2026,
	title = {{PAOFLOW}: an automated suite for ab initio electronic, transport, and topological properties of materials},
	author = {Anooja Jayaraj and Sergio Alvarruiz and Mia Falatko and Zhiren He and Jonathan Red and Caua F. Schuch and Karma Tenzin and Chen Ye, Chao and Davide Ceresoli and Marcio Costa and Stefano Curtarolo and Jagoda S{\l}awi{\'n}ska and  Buongiorno Nardelli, Marco},
	journal = {arXiv:2609.21984},
	year = {2026},
}

@article{z2pack_2017,
  title = {{Z2Pack: Numerical implementation of hybrid Wannier centers for identifying topological materials}},
  author = {Gresch, Dominik and Aut\`es, Gabriel and Yazyev, Oleg V. and Troyer, Matthias and Vanderbilt, David and Bernevig, B. Andrei and Soluyanov, Alexey A.},
  journal = {Phys. Rev. B},
  volume = {95},
  issue = {7},
  pages = {075146},
  numpages = {24},
  year = {2017},
  month = {Feb},
  publisher = {American Physical Society},
  doi = {10.1103/PhysRevB.95.075146}
}

@article{z2pack_2011,
  title = {Computing topological invariants without inversion symmetry},
  author = {Soluyanov, Alexey A. and Vanderbilt, David},
  journal = {Phys. Rev. B},
  volume = {83},
  issue = {23},
  pages = {235401},
  numpages = {9},
  year = {2011},
  month = {Jun},
  publisher = {American Physical Society},
  doi = {10.1103/PhysRevB.83.235401}
}

@article{wanniertools_2018,
title = "WannierTools : An open-source software package for novel topological materials",
journal = "Comput. Phys. Commun.",
volume = "224",
pages = "405 - 416",
year = "2018",
issn = "0010-4655",
doi = "10.1016/j.cpc.2017.09.033",
author = "QuanSheng Wu and ShengNan Zhang and Hai-Feng Song and Matthias Troyer and Alexey A. Soluyanov"
}

@article{kane_mele_2005,
	title = {$\mathbb{Z}_{2}$ Topological Order and the Quantum Spin {Hall} Effect},
	author = {Kane, C. L. and Mele, E. J.},
	journal = {Phys. Rev. Lett.},
	volume = {95},
	issue = {14},
	pages = {146802},
	year = {2005},
	month = {Sep},
	publisher = {American Physical Society},
	doi = {10.1103/PhysRevLett.95.146802},
}

@article{bernevig_2006,
	title = {Quantum Spin {Hall} Effect and Topological Phase Transition in {HgTe} Quantum Wells},
	author = {Bernevig, B. Andrei and Hughes, Taylor L. and Zhang, Shou-Cheng},
	journal = {Science},
	volume = {314},
	number = {5806},
	pages = {1757--1761},
	year = {2006},
	doi = {10.1126/science.1133734},
}

@article{konig_2007,
	title = {Quantum Spin {Hall} Insulator State in {HgTe} Quantum Wells},
	author = {K{\"o}nig, Markus and Wiedmann, Steffen and Br{\"u}ne, Christoph and Roth, Andreas and Buhmann, Hartmut and Molenkamp, Laurens W. and Qi, Xiao-Liang and Zhang, Shou-Cheng},
	journal = {Science},
	volume = {318},
	number = {5851},
	pages = {766--770},
	year = {2007},
	doi = {10.1126/science.1148047},
}

@article{sheng_2005,
	title = {Nondissipative Spin {Hall} Effect via Quantized Edge Transport},
	author = {Sheng, L. and Sheng, D. N. and Ting, C. S. and Haldane, F. D. M.},
	journal = {Phys. Rev. Lett.},
	volume = {95},
	issue = {13},
	pages = {136602},
	year = {2005},
	month = {Sep},
	publisher = {American Physical Society},
	doi = {10.1103/PhysRevLett.95.136602},
}

@article{sheng_2006,
	title = {Quantum Spin-{Hall} Effect and Topologically Invariant {Chern} Numbers},
	author = {Sheng, D. N. and Weng, Z. Y. and Sheng, L. and Haldane, F. D. M.},
	journal = {Phys. Rev. Lett.},
	volume = {97},
	issue = {3},
	pages = {036808},
	year = {2006},
	month = {Jul},
	publisher = {American Physical Society},
	doi = {10.1103/PhysRevLett.97.036808},
}

@article{prodan_2009,
	title = {Robustness of the spin-{Chern} number},
	author = {Prodan, Emil},
	journal = {Phys. Rev. B},
	volume = {80},
	issue = {12},
	pages = {125327},
	year = {2009},
	month = {Sep},
	publisher = {American Physical Society},
	doi = {10.1103/PhysRevB.80.125327},
}

@article{yang_2011,
	title = {Time-Reversal-Symmetry-Broken Quantum Spin {Hall} Effect},
	author = {Yang, Yunyou and Xu, Zhong and Sheng, L. and Wang, Baigeng and Xing, D. Y. and Sheng, D. N.},
	journal = {Phys. Rev. Lett.},
	volume = {107},
	issue = {6},
	pages = {066602},
	year = {2011},
	month = {Aug},
	publisher = {American Physical Society},
	doi = {10.1103/PhysRevLett.107.066602},
}

@article{ezawa_2013,
	title = {High Spin-{Chern} Insulators with Magnetic Order},
	author = {Ezawa, Motohiko},
	journal = {Sci. Rep.},
	volume = {3},
	pages = {3435},
	year = {2013},
	doi = {10.1038/srep03435},
}

@article{roadmap_2024,
	title = {{2024 roadmap on 2D topological insulators}},
	author = {Bent Weber and Michael S. Fuhrer and Xian-Lei Sheng and Shengyuan A. Yang and Ronny Thomale and Saquib Shamim and Laurens W. Molenkamp and David Cobden and Dmytro Pesin and Harold J. W. Zandvliet and Pantelis Bampoulis and Ralph Claessen and Fabian R. Menges and Johannes Gooth and Claudia Felser and Chandra Shekhar and Anton Tadich and Mengting Zhao and Mark T. Edmonds and Junxiang Jia and Maciej Bieniek and Jukka I. V{\"a}yrynen and Dimitrie Culcer and Bhaskaran Muralidharan and Muhammad Nadeem},
	journal = {J. Phys. Mater.},
	volume = {7},
	number = {2},
	pages = {022501},
	year = {2024},
	doi = {10.1088/2515-7639/ad2083},
}

@article{wang_exfoliation_2023,
	title = {Scalable high yield exfoliation for monolayer nanosheets},
	author = {Wang, Zhuyuan and Yan, Xue and Hou, Qinfu and Liu, Yue and Zeng, Xiangkang and Kang, Yuan and Zhao, Wang and Li, Xuefeng and Yuan, Shi and Qiu, Ruosang and Uddin, Md Hemayet and Wang, Ruoxin and Xia, Yun and Jian, Meipeng and Kang, Yan and Gao, Li and Liang, Songmiao and Liu, Jefferson Zhe and Wang, Huanting and Zhang, Xiwang},
	journal = {Nat. Commun.},
	volume = {14},
	pages = {236},
	year = {2023},
	doi = {10.1038/s41467-022-35569-8},
}

@article{tajkov_2023,
	title = {Mechanical strain induced topological phase changes of few layer {ZrTe}$_5$},
	author = {Tajkov, Zolt{\'a}n and Kandrai, Konr{\'a}d and Nagy, D{\'a}niel and Tapaszt{\'o}, Levente and Koltai, J{\'a}nos and Nemes-Incze, P{\'e}ter},
	journal = {arXiv:2311.04721v1},
	year = {2023},
}

@article{bai_2022,
	title = {{Doubled quantum spin Hall effect with high-spin Chern number in $\alpha$-antimonene and $\alpha$-bismuthene}},
	author = {Bai, Yingxi and Cai, Linke and Mao, Ning and Li, Runhan and Dai, Ying and Huang, Baibiao and Niu, Chengwang},
	journal = {Phys. Rev. B},
	volume = {105},
	issue = {19},
	pages = {195142},
	year = {2022},
	month = {May},
	publisher = {American Physical Society},
	doi = {10.1103/PhysRevB.105.195142},
}

@article{liu_2024,
	title = {{Orbital doublet driven even-spin Chern insulators}},
	author = {Liu, Lu and Liu, Yuntian and Li, Jiayu and Wu, Hua and Liu, Qihang},
	journal = {Phys. Rev. B},
	volume = {110},
	issue = {3},
	pages = {035161},
	year = {2024},
	month = {Jul},
	publisher = {American Physical Society},
	doi = {10.1103/PhysRevB.110.035161},
}

@article{chen_2017,
	title = {Spectroscopic evidence for bulk-band inversion and three-dimensional massive {Dirac} fermions in {ZrTe}$_5$},
	author = {Chen, Zhi-Guo and Chen, R. Y. and Zhong, R. D. and Schneeloch, John and Zhang, C. and Huang, Y. and Qu, Fanming and Yu, Rui and Li, Q. and Gu, G. D. and Wang, N. L.},
	journal = {Proc. Natl. Acad. Sci. U.S.A.},
	volume = {114},
	number = {5},
	pages = {816--821},
	year = {2017},
	doi = {10.1073/pnas.1613110114},
}

@article{mutch_2019,
	title = {Evidence for a strain-tuned topological phase transition in {ZrTe}$_5$},
	author = {Mutch, Joshua and Chen, Wei-Chih and Went, Preston and Qian, Tiema and Wilson, Ilham Zaky and Andreev, Anton and Chen, Cheng-Chien and Chu, Jiun-Haw},
	journal = {Sci. Adv.},
	volume = {5},
	number = {8},
	pages = {eaav9771},
	year = {2019},
	doi = {10.1126/sciadv.aav9771},
}

@article{xu_2018,
	title = {Temperature-Driven Topological Phase Transition and Intermediate {Dirac} Semimetal Phase in {ZrTe}$_5$},
	author = {Xu, B. and Zhao, L. X. and Marsik, P. and Sheveleva, E. and Lyzwa, F. and Dai, Y. M. and Chen, G. F. and Qiu, X. G. and Bernhard, C.},
	journal = {Phys. Rev. Lett.},
	volume = {121},
	issue = {18},
	pages = {187401},
	year = {2018},
	month = {Nov},
	publisher = {American Physical Society},
	doi = {10.1103/PhysRevLett.121.187401},
}

@article{zhang_2021,
	title = {Observation and control of the weak topological insulator state in {ZrTe}$_5$},
	author = {Zhang, Peng and Noguchi, Ryo and Kuroda, Kenta and Lin, Chun and Kawaguchi, Kaishu and Yaji, Koichiro and Harasawa, Ayumi and Lippmaa, Mikk and Nie, Simin and Weng, Hongming and Kandyba, V. and Giampietri, A. and Barinov, A. and Li, Qiang and Gu, G. D. and Shin, Shik and Kondo, Takeshi},
	journal = {Nat. Commun.},
	volume = {12},
	number = {1},
	pages = {406},
	year = {2021},
	doi = {10.1038/s41467-020-20564-8},
}

@article{wang_2021,
	title = {Magneto-transport evidence for strong topological insulator phase in {ZrTe}$_5$},
	author = {Wang, Jingyue and Jiang, Yuxuan and Zhao, Tianhao and Dun, Zhiling and Miettinen, Anna L. and Wu, Xiaosong and Mourigal, Martin and Zhou, Haidong and Pan, Wei and Smirnov, Dmitry and Jiang, Zhigang},
	journal = {Nat. Commun.},
	volume = {12},
	number = {1},
	pages = {6758},
	year = {2021},
	doi = {10.1038/s41467-021-27119-5},
}

@article{tang_2021,
	title = {Two-Dimensional Quantum {Hall} Effect and Zero Energy State in Few-Layer {ZrTe}$_5$},
	author = {Tang, Fangdong and Wang, Peipei and He, Mingquan and Isobe, Masahiko and Gu, Genda and Li, Qiang and Zhang, Liyuan and Smet, Jurgen H.},
	journal = {Nano Lett.},
	volume = {21},
	number = {14},
	pages = {5998--6004},
	year = {2021},
	month = {Jul},
	doi = {10.1021/acs.nanolett.1c00958},
}

@article{xu_2024,
	title = {Realization of monolayer {ZrTe}$_5$ topological insulators with wide band gaps},
	author = {Xu, Yong-Jie and Cao, Guohua and Li, Qi-Yuan and Xue, Cheng-Long and Zhao, Wei-Min and Wang, Qi-Wei and Dou, Li-Guo and Du, Xuan and Meng, Yu-Xin and Wang, Yuan-Kun and Gao, Yu-Hang and Jia, Zhen-Yu and Li, Wei and Ji, Lianlian and Li, Fang-Sen and Zhang, Zhenyu and Cui, Ping and Xing, Dingyu and Li, Shao-Chun},
	journal = {Nat. Commun.},
	volume = {15},
	number = {1},
	pages = {4784},
	year = {2024},
	doi = {10.1038/s41467-024-49197-x},
}

@article{qi_2011,
	title = {Topological insulators and superconductors},
	author = {Qi, Xiao-Liang and Zhang, Shou-Cheng},
	journal = {Rev. Mod. Phys.},
	volume = {83},
	issue = {4},
	pages = {1057--1110},
	year = {2011},
	month = {Oct},
	publisher = {American Physical Society},
	doi = {10.1103/RevModPhys.83.1057},
}

@article{tang_2019,
        author = {Fangdong Tang and Yafei Ren and Peipei Wang and Ruidan Zhong and John Schneeloch and Shengyuan A. Yang and Kun Yang and Patrick A. Lee and Genda Gu and Zhenhua Qiao and Liyuan Zhang},
	title = "{Three-dimensional quantum Hall effect and metal–insulator transition in ZrTe$_5$}",
	journal = {Nature},
	volume = {569},
	number = {7757},
	pages = {537--541},
	year = {2019},
	month = {May},
	doi = {10.1038/s41586-019-1180-9},
}

@article{galeski_2021,
	title = "{Origin of the quasi-quantized Hall effect in ZrTe$_{5}$}",
	  author    = {S. Galeski and T. Ehmcke and R. Wawrzyńczak and P. M. Lozano and K. Cho and A. Sharma and S. Das and F. Küster and P. Sessi and M. Brando and R. Küchler and A. Markou and M. König and P. Swekis and C. Felser and Y. Sassa and Q. Li and G. Gu and M. V. Zimmermann and O. Ivashko and D. I. Gorbunov and S. Zherlitsyn and T. Förster and S. S. P. Parkin and J. Wosnitza and T. Meng and J. Gooth},
	journal = {Nat. Commun.},
	volume = {12},
	number = {1},
	pages = {3197},
	year = {2021},
	month = {May},
	doi = {10.1038/s41467-021-23435-y},
}

@article{weng_2014,
	title = "{Transition-metal pentatelluride $\mathrm{ZrTe}_{5}$ and $\mathrm{HfTe}_{5}$: a paradigm for large-gap quantum spin Hall insulators}",
	author = {Weng, Hongming and Dai, Xi and Fang, Zhong},
	journal = {Phys. Rev. X},
	volume = {4},
	issue = {1},
	pages = {011002},
	numpages = {8},
	year = {2014},
	month = {Jan},
	publisher = {American Physical Society},
	doi = {10.1103/PhysRevX.4.011002}
}

@article{fan_2017,
	title="{Transition between strong and weak topological insulator in ZrTe$_5$ and HfTe$_5$}",
    author  = {Zongjian Fan and Qi-Feng Liang and Y. B. Chen and Shu-Hua Yao and Jian Zhou},
	journal={Sci. Rep.},
	volume={7},
	pages={45667},
	year={2017},
	month = {Abr},
	doi={10.1038/srep45667}
}

@article{chenye_2025,
  title = {First-principles studies of fermiology in topological phases of bulk {ZrTe}$_{5}$},
  author = {Chen Ye, Chao and Kreminska, Yuliia and Ye, Jianting and S{\l}awi{\'n}ska, Jagoda},
  journal = {Phys. Rev. Materials},
  volume = {9},
  issue = {5},
  pages = {054204},
  numpages = {9},
  year = {2025},
  month = {May},
  publisher = {American Physical Society},
  doi = {10.1103/PhysRevMaterials.9.054204},
}

@article{manzoni_2016,
	title = "{Evidence for a strong topological insulator phase in ${\mathrm{ZrTe}}_{5}$}",
	  author = {Manzoni, G. and Gragnaniello, L. and Aut\`es, G. and Kuhn, T. and Sterzi, A. and Cilento, F. and Zacchigna, M. and Enenkel, V. and Vobornik, I. and Barba, L. and Bisti, F. and Bugnon, Ph. and Magrez, A. and Strocov, V. N. and Berger, H. and Yazyev, O. V. and Fonin, M. and Parmigiani, F. and Crepaldi, A.},
	journal = {Phys. Rev. Lett.},
	volume = {117},
	issue = {23},
	pages = {237601},
	numpages = {5},
	year = {2016},
	month = {Nov},
	publisher = {American Physical Society},
	doi = {10.1103/PhysRevLett.117.237601}
}

@article{tajkov_2022,
	author = {Zoltán Tajkov and Dániel Nagy and Konrád Kandrai and János Koltai and László Oroszlány and Péter Süle and Zsolt E. Horváth and Péter Vancsó and Levente Tapasztó and Péter Nemes-Incze},
	title = "{Revealing the topological phase diagram of ${\mathrm{ZrTe}}_{5}$ using the complex strain fields of microbubbles}",
	journal = {npj Comput. Mater.},
	year = {2022},
	volume = {8},
	number = {1},
	pages = {177},
	doi = {10.1038/s41524-022-00854-z},
	issn = {2057-3960}
}

@article{qian_2014,
author = {Xiaofeng Qian  and Junwei Liu  and Liang Fu  and Ju Li },
title = {Quantum spin {Hall} effect in two-dimensional transition metal dichalcogenides},
journal = {Science},
volume = {346},
number = {6215},
pages = {1344-1347},
year = {2014},
doi = {10.1126/science.1256815},
}

@article{hse_2009,
  title = {{Heyd-Scuseria-Ernzerhof hybrid functional for calculating the lattice dynamics of semiconductors}},
  author = {Hummer, Kerstin and Harl, Judith and Kresse, Georg},
  journal = {Phys. Rev. B},
  volume = {80},
  issue = {11},
  pages = {115205},
  numpages = {12},
  year = {2009},
  month = {Sep},
  publisher = {American Physical Society},
  doi = {10.1103/PhysRevB.80.115205},
}

@article{roy_2022,
  title = {Unconventional spin {Hall} effects in nonmagnetic solids},
  author = {Roy, Arunesh and Guimar{\~a}es, Marcos H. D. and S{\l}awi{\'n}ska, Jagoda},
  journal = {Phys. Rev. Materials},
  volume = {6},
  issue = {4},
  pages = {045004},
  numpages = {10},
  year = {2022},
  month = {Apr},
  publisher = {American Physical Society},
  doi = {10.1103/PhysRevMaterials.6.045004},
}

@article{tan_2025,
title = {Stacking-induced type-II quantum spin Hall insulators with high spin Chern number in unconventional magnetism},
journal = {Science Bulletin},
volume = {71},
number = {9},
pages = {2196-2199},
year = {2026},
author = {Chao-Yang Tan and Panjun Feng and Ze-Feng Gao and Fengjie Ma and Peng-Jie Guo and Zhong-Yi Lu},
}

@article{stuhler_2022,
  author = {St\"{u}hler, Reinald and Kowalewski, Artur and Reis, Felix and Jungblut, David and Dominguez, Fernando and Scharf, Benedikt and Li, Gang and Sch\"{a}fer, J\"{o}rg and Hankiewicz, Ewelina M. and Claessen, Ralph},
  title = {Effective lifting of the topological protection of quantum spin {Hall} edge states by edge coupling},
  journal = {Nat. Commun.},
  year = {2022},
  volume = {13},
  number = {1},
  pages = {3480},
  doi = {10.1038/s41467-022-30996-z},
}

@article{Kang2024,
  title={{Double quantum spin Hall phase in moir{\'e} WSe$_2$}},
  author={Kang, Kaifei and Qiu, Yichen and Watanabe, Kenji and Taniguchi, Takashi and Shan, Jie and Mak, Kin Fai},
  journal={Nano Lett.},
  volume={24},
  number={46},
  pages={14901--14907},
  year={2024},
  publisher={ACS Publications},
  doi = {10.1021/acs.nanolett.4c05308}
}

@article{kang_2024,
  title={{Evidence of the fractional quantum spin Hall effect in moir{\'e} MoTe$_2$}},
  author={Kang, Kaifei and Shen, Bowen and Qiu, Yichen and Zeng, Yihang and Xia, Zhengchao and Watanabe, Kenji and Taniguchi, Takashi and Shan, Jie and Mak, Kin Fai},
  journal={Nature},
  volume={628},
  number={8008},
  pages={522--526},
  year={2024},
  publisher={Nature Publishing Group UK London}
}

@article{qiu_2016,
  title={{Observation of optical and electrical in-plane anisotropy in high-mobility few-layer ZrTe$_5$}},
  author={Qiu, Gang and Du, Yuchen and Charnas, Adam and Zhou, Hong and Jin, Shengyu and Luo, Zhe and Zemlyanov, Dmitry Y and Xu, Xianfan and Cheng, Gary J and Ye, Peide D},
  journal={Nano Lett.},
  volume={16},
  number={12},
  pages={7364--7369},
  year={2016},
  publisher={ACS Publications}
}

@article{coleman_2011,
  title   = {Two-Dimensional Nanosheets Produced by Liquid Exfoliation of Layered Materials},
  author  = {Coleman, Jonathan N. and Lotya, Mustafa and O'Neill, Arlene and Bergin, Shane D. and King, Paul J. and Khan, Umar and Young, Karen and Gaucher, Alexandre and De, Sukanta and Smith, Ronan J. and Shvets, Igor V. and Arora, Sunil K. and Stanton, George and Kim, Hye Young and Lee, Kangho and Kim, Gyu Tae and Duesberg, Georg S. and Hallam, Toby and Boland, John J. and Wang, Jing Jing and Donegan, John F. and Grunlan, Jaime C. and Moriarty, Gregory and Shmeliov, Aleksey and Nicholls, Rebecca J. and Perkins, James M. and Grieveson, Eleanor M. and Theuwissen, Koenraad and McComb, David W. and Nellist, Peter D. and Nicolosi, Valeria},
  journal = {Science},
  volume  = {331},
  number  = {6017},
  pages   = {568--571},
  year    = {2011},
  doi     = {10.1126/science.1194975}
}

@article{backes_2020,
  title   = {Equipartition of Energy Defines the Size–Thickness Relationship in Liquid-Exfoliated Nanosheets},
  author  = {Backes, Claudia and Campi, Davide and Szydlowska, Beata M. and Synnatschke, Kevin and Ojala, Ezgi and Rashvand, Farnia and Harvey, Andrew and Griffin, Aideen and Sofer, Zdenek and Marzari, Nicola and O'Regan, David D. and Coleman, Jonathan N.},
  journal = {ACS Nano},
  year    = {2020},
  doi     = {10.1021/acsnano.0c01333}
}

@article{hernandez_2008,
  title   = {High-yield production of graphene by liquid-phase exfoliation of graphite},
  author  = {Hernandez, Yenny and Nicolosi, Valeria and Lotya, Mustafa and Blighe, Fiona M. and Sun, Zhenyu and De, Sukanta and McGovern, Ian T. and Holland, Brendan and Byrne, Mike and Gun'ko, Yurii K. and Boland, John J. and Niraj, P. and Duesberg, Georg S. and Krishnamurthy, S. and Goodhue, Rory and Hutchison, John and Scardaci, Vincenzo and Ferrari, Andrea C. and Coleman, Jonathan N.},
  journal = {Nature Nanotechnology},
  volume  = {3},
  pages   = {563--568},
  year    = {2008},
  doi     = {10.1038/nnano.2008.215}
}

\begin{figure*} 
	\centering
	\includegraphics[width=0.8\textwidth]{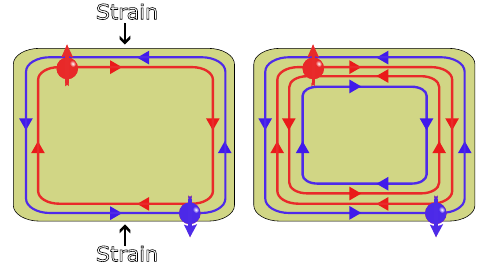}
	%
	\label{fig:toc}
\end{figure*}

\end{document}